\documentclass[prl,twocolumn,groupedaddress]{revtex4}
\begin{document}
\title[]{Comment on ``Freezing by heating in a driven mesoscopic system''}
\author{Leo Radzihovsky}
\author{Noel A. Clark}
\affiliation{Department of Physics, University of Colorado,
   Boulder, CO 80309}

\date{\today}

\begin{abstract}
%  We point out that the phenomenon ``heating by freezing'' discovered
%  in simulations by Helbing, et al. extend to equilibrium
%  systems as well. We argue that such reentrant freezing can, for
%  example, be realized in two-dimensional colloidal systems
%  subjected to a one-dimensional periodic potential.
\end{abstract}

\maketitle 
%\vspace{-1.5cm} 

In a recent Letter Helbing et al.\cite{Helbing} found a fascinating
fluctuation-induced ordering, which they dubbed  "freezing by heating", in a
nonequilibrium model system of particles confined to a 2d narrow strip and
driven to move in opposite directions: counter flowing chains of particles jam
up and crystallize when sufficiently large noise is introduced into the
system.    In this Comment we point out that this kind of fluctuation-induced
ordering is {\em not} an intrinsically {\em nonequilibrium} effect, and that
it can be expected to occur in a variety of {\em equilibrium} systems, with
reentrant order appearing as a function of temperature.  We argue that this is
a consequence of a familiar Helfrich-like interaction that had been used to
understand a variety of phenomena ranging from the lamellar phase of
fluctuating membranes to magnetization curves of a vortex liquid state of type
II superconductors.

We demonstrate the mechanism behind reentrant melting with an equilibrium
model of 2d colloids subjected to a 1d periodic potential, that received
considerable attention recently due to the experimental discovery of {\em
isothermal} melting that is reentrant as a function of strength of the
periodic potential (produced by two interfering laser beams) \cite{Wei}, the
so-called reentrant ``laser-induced melting'' (RLIM).\cite{LIFclark}

\begin{figure}
\centering
\setlength{\unitlength}{1mm}
\begin{picture}(150,25)(0,0)
\put(-20,-50){\begin{picture}(150,0)(0,0)
\includegraphics{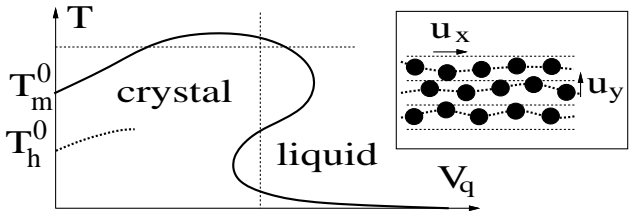}
\end{picture}}
\end{picture}
\caption{Schematic phase diagram of colloidal system in the inset showing 
  the mechanism for fluctuation driven freezing.}
\label{phasediagram}
\end{figure}

Since at all $T$ the orientational symmetry and translational symmetry
along $y$ are {\em explicitly} broken by the 1d periodic potential
with troughs running along $x$ and spacing $a=2\pi/q$, the only
remaining feature distinguishing a crystal from the modulated liquid
is the spontaneous translational order along the troughs,
characterized by quasi-long ranged $u_x$
phonon correlations and a finite shear modulus $\mu_{xy}$.

The mechanism behind RLIM can be understood as follows\cite{RFN}. 
Just above melting temperature $T_m(V_q)$, the dominant effect of
increasing a {\em weak} periodic potential $V_q$ is to suppress
thermal fluctuations, thereby leading to LIF\cite{LIFclark}. In
contrast, for {\em large} $V_{q}$, the suppression of the
out-of-trough $u_y$ fluctuations has the dominant effect of reducing
the inter-row interactions, thereby decoupling the particles in the
neighboring rows, reducing the effective shear modulus
$\mu_{xy}(T,V_{q})$, and leading to RLIM.\cite{Wei}

Since it is clearly the ratio $T/V_q$ that determines the extent of
the fluctuation-driven inter-row interaction, it follows that at
large {\em fixed} $V_q$ increasing $T$ will increase the inter-row
interaction and the modulus $\mu_{xy}(T,V_{q})$. Because $T_m$ is
typically a monotonically increasing function of the shear
modulus\cite{RFN} thermal enhancement of shear modulus implies an
equilibrium analog of the ``freezing by heating'' transition
discovered by Helbing, et al.\cite{Helbing}.

Above qualitative argument can be formalized by an explicit
calculation.  Although $T$ dependence of both the bulk and the shear
moduli maybe important and can be studied in full detail\cite{RFN},
in order to illustrate the effect, it is sufficient to focus on the
effective shear modulus $\mu_{xy}(T,V_q)$.  Clearly this modulus is
proportional to the interaction potential between nearest particles
in adjacent rows, typically of Debye-screened form $U_D\approx
e^{-(a+u_y)/\xi}$ (though only its short-range nature is important),
where $\xi$ is the the screening length. Since phonon fluctuations
$u_y$ transverse to the troughs are ``optical'' (i.e., ``massive'',
as opposed to the acoustic $u_x$ phonon) we can average over fast
$u_y$ fluctuations obtaining an effective shear modulus
$\mu_{xy}(T,V_q)\approx \mu_0 e^{\langle
u_y^2\rangle/2\xi^2}\rangle\approx \mu_0 e^{T/V_q}$ that clearly
increases with $T$.\cite{comment} We can then compute $\langle
u_x^2\rangle\approx T/\mu_{xy}(T,V_q)\approx T e^{-T/V_q}/\mu_0$ and
use Lindemann criterion $\langle u_x^2\rangle=d^2$ to estimate
$T_m(V_q)$. This clearly gives two roots for $T_m$, with the lower
corresponding to the conventional melting and the upper to the
freezing by heating (see figure). At sufficiently high $T$, we expect
above elastic calculation to become inaccurate and for the
fluctuation enhancement of $\mu_{xy}(T,V_q)$ to saturate, leading to
re-melting into a modulated liquid.

While extending above analysis to the nonequilibrium case studied by
Helbing, et al.\cite{Helbing} remains an interesting and challenging
problem, we believe that fluctuation-driven reentrant freezing
mechanism described above is also at play in their system.

Support by the NSF MRSEC DMR-0213918 and the Packard Foundation (LR).

\end{document}